\documentclass[seceq,supplement]{ptptex}

\usepackage{graphicx}
\title{%        %You can use \\ for explicit line-break.
Resonant states of open quantum systems%
}

\author{%       %Use \scshape for the family name.
\textsc{Naomichi Hatano}%$^1$ and Gonzalo \textsc{Ordonez}$^2$%
}

\inst{%     %Affiliation, neglected when [addenda] or [errata].
Institute of Industrial Science, University of Tokyo\\
Komaba 4-6-1, Meguro, Tokyo 153-8505%\\
}

\abst{%         %This abstract is neglected when [addenda] or [errata].
We first show that quantum resonant states observe particle number conservation and hence are consistent with the probabilistic interpretation of quantum mechanics.
We then present for a class of quantum open systems, a resonant-state expansion of the sum of the retarded and advanced Green's functions.
The expansion is given purely in terms of all discrete eigenstates and does not contain any background integrals.
Using the expansion, we argue that the Fano asymmetry of resonance peaks is interpreted as interference between discrete eigenstates.
We microscopically derive the Fano parameters for several cases.
}

\begin{document}

\maketitle

\section{Introduction}
In the present paper, we review recent results on quantum resonant states in one-dimensional systems.
Quantum resonant states have been of interest for many decades in various branches of physics~\cite{Newton82,Kukulin89,Brandas90,Razavy03}.
Researches have been carried out perhaps primarily in the fields of nuclear physics~\cite{Bethe35,Bethe36,Hulthen42a,Hulthen42b,Jost47,Wigner55,Nakanishi58,Humblet61,Rosenfeld61,Humblet62,Humblet64a,Jeukenne64,Humblet64b,Mahaux65,Rosenfeld65,GarciaCalderon76,GarciaCalderon07a,GarciaCalderon07b,Bohm89,Homma97,Myo97,Myo98,Masui99,Suzuki05,Aoyama06,Amrein87,Carvalho02,Ahmed04,Kelkar04,Jain05,Amrein06,Rotter09} as well as atomic and molecular physics~\cite{Smith62,Burke62,Schulz64,Ho77,Agostini79,Freeman87,Moiseyev90,Peskin92,Peskin93,Moiseyev97,Brisker08,Moiseyev08}.
Thanks to the development of nanotechnology in recent years, quantum resonant states are also of current interest in condensed-matter physics~\cite{Tekman93,Gores00,Zacharia01,Clerk01,Racec01,Kang01,Torio02,Kobayashi02,Kobayashi03,Kobayashi04,Sato05,Kim03,Babic04,Lu05,Chakrabarti06,Franco06,Joe07,Fujimoto08}.

\begin{figure}
\centering
\includegraphics[width=0.75\textwidth]{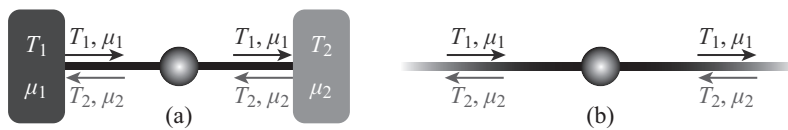}
\caption{(a) An experimental setup of two-terminal measurement of the conductance of a mesoscopic system.
The mesoscopic system is connected to finite leads, at whose ends are two heat baths with the temperature and the chemical potential $(T_1,\mu_1)$ and $(T_2,\mu_2)$, respectively.
(b) A theoretical setup of calculating the conductance of a mesoscopic system.
The mesoscopic system is connected to semi-infinite leads.}
\label{fig0}
\end{figure}

Electronic properties of mesoscopic systems such as quantum dots may be measured in an experimental setup shown in Fig.~\ref{fig0}(a).
(There are, of course, more complicated setups such as four-terminal measurement, but for the moment, let us consider only the two-terminal measurement for simplicity.)
The celebrated Landauer formula~\cite{Datta95} tells us that the conductance is given by the transmission coefficient of the quantum scattering problem solved in a theoretical setup shown in Fig.~\ref{fig0}(b).
(We here suppose that electrons do not interact with each other on the leads.)

The essential reason why the conductance measured in the setup of Fig.~\ref{fig0}(a) can be given by the quantum scattering problem solved in the setup of Fig.~\ref{fig0}(b) may be as follows.
In the setup of Fig.~\ref{fig0}(a), an electron that comes from left into the central system is in equilibrium with the left heat bath.
After scattered by the system, the electron, if transmitted, comes out onto the right heat bath.
It never comes back into the system again (at least, not coherently), because there are an infinite number of degrees of freedom in the right heat bath.
It is important to notice this property of a heat bath that it absorbs an electron and never emits it again coherently.
This property is kept in the setup of Fig.~\ref{fig0}(b).
An electron that comes from the left infinity, if transmitted, comes out of the system into the right infinity and never comes back.
This is why the conductance measured in the setup of Fig.~\ref{fig0}(a) can be given by the transmission coefficient calculated in the setup of Fig.~\ref{fig0}(b).

In other words, nanotechnology lets us consider a pure quantum scattering problem in the field of condensed-matter physics, if electronic interactions are negligible in the leads.
It is in this spirit that we have studied quantum resonant states in the setup of Fig.~\ref{fig0}(b)~\cite{Sasada05,Nishino07,Sasada08,Hatano08,Hatano09,Sasada09,Nishino09,Imamura09}.
In the present paper, we review specifically two of our recent studies in this direction~\cite{Hatano08,Hatano09,Sasada09}.
In one, we revealed physical aspects of quantum resonant states.
In particular, we discovered that a resonant state conserves the particle number and hence is consistent with the probabilistic interpretation of quantum mechanics, in contrast to conventional beliefs.
In the other study, we discovered a resonant-state expansion of the Green's functions.
We showed for a specific class of systems that the sum of the retarded and advanced Green's functions is expanded in terms of all discrete eigenstates including the resonant, anti-resonant, bound and anti-bound states and it has no background integrals.
With the use of the expansion, we discussed that the Fano asymmetry of the resonance peaks is originated in interference between various discrete eigenstates.
In particular, we found a type of asymmetry that was probably not considered in Fano's original argument~\cite{Fano61}.
(Throughout the present paper, we consider one-particle problems.
See, for example, Refs.~\citen{Nishino09,Imamura09} for the cases where electrons interact with each other on the dot.)

The paper is organized as follows.
In Section~\ref{sec2}, we will review our study on physical aspects of the quantum resonant state, particularly its particle-number conservation.
Section~\ref{sec3} will present derivation of the resonant-state expansion of the sum of the retarded and advanced Green's functions.
Using the expansion, we will consider the Fano asymmetry of the resonance peak in Section~\ref{sec4}.
Several types of the Fano parameter will be given.
The final section will be devoted to a summary.

\section{Probabilistic interpretation of resonant states}
\label{sec2}

\subsection{Dynamical and static views of quantum resonant states}

In the present section, we will discuss physical pictures of resonant states.
Before going into details, let us make explicit that there are two views of quantum resonant phenomena; namely, a dynamical view and a static view.
In a dynamical view, quantum resonance is often described in the following way.
A quantum particle, or a wave packet comes into a trapping potential, stays there for a while, bouncing back and forth in the trapping potential, before it comes out.
The typical time duration of the particle's stay in the trapping potential is the resonance lifetime.
The lifetime can be long when the energy expectation value of the incoming wave packet is close to, or ``resonates with," a quasi-bound state of the trapping potential.

In the present paper, on the other hand, we will mainly consider the quantum resonance in the framework of its static view.
As we will discuss below, the resonance in the static view is an eigenstate of the standard time-independent Schr\"{o}dinger equation but under a non-standard boundary condition. %, namely the Siegert condition.
We will show that the openness of the system generates non-Hermiticity of the Hamiltonian and thereby produces eigenstates with complex eigenvalues.
The resonance lifetime is given by the reciprocal of the imaginary part of the complex eigenvalue.
The resonant state is a stationary function multiplied by a decay factor that makes the wave amplitude in the trapping potential decrease exponentially from infinite past to infinite future;
the decay rate is the resonance lifetime.

The static view of resonance appears to be less popular than the dynamical view of resonance, perhaps because of its seemingly unphysical features.
One feature that is not accepted well is the exponential decay of the wave amplitude, which appears to indicate breaking of the particle-number conservation.
Another feature that is often found problematic is the fact that the wave function of the resonant eigenstate diverges in space far away from the trapping potential.

The main purpose of the present section is to review a recent discussion on the particle-number conservation of the resonant state~\cite{Hatano08}.
The discussion, in fact, relates the above two ``unphysical" features and sublimates them into a unified physical picture of the resonant state as an eigenstate.
The exponential decay is shown to be caused by leak of the particles from the trapping potential.
This leak then manifests itself in the form of the spatial divergence of the wave function.

\subsection{Direct definition of resonant states in the static view}

In the present subsection, we will review two types of definition of resonant states in the static view.
We could call them an indirect definition and a direct one.
The indirect definition may be the one that we can find in many textbooks.
In the present paper, however, we will mainly use the direct definition.
Before proceeding with the direct definition, let us make sure that the two types of definition are equivalent.

\begin{figure}
\centering
\includegraphics[width=0.55\textwidth]{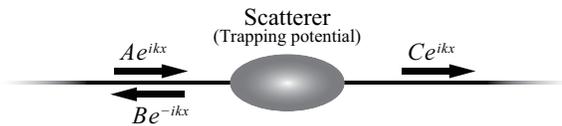}
\caption{A quantum scatterer in one dimension.}
\label{fig1}
\end{figure}

In the indirect definition, a resonant state is a pole of the $S$ matrix.
For simplicity, let us work in one dimension hereafter and consider the situation in Fig.~\ref{fig1}, where we have a trapping potential with a compact support.
Then the $S$ matrix is given by the transmission and reflection amplitudes.
More specifically, we consider an incident wave $Ae^{ikx}$ and its reflection wave $Be^{-ikx}$ as well as its transmission wave $Ce^{ikx}$.
In solving the Schr\"{o}dinger equation, we require the wave to be continuous and differentiable around the scatterer.
This produces in the amplitudes the $k$ dependence, $A(k)$, $B(k)$ and $C(k)$, and through the dispersion relation $E=\hbar^2 k^2/2m$, the $E$ dependence.
The reflection and transmission amplitudes are given, respectively, by
%\begin{align}\label{eq2-10}
$r(k)\equiv B(k)/A(k)$
%\qquad\mbox{and}\qquad
and $t(k)\equiv C(k)/A(k)$.
%\end{align}
We then realize that the zeros of $A(k)$ give the poles of $r(k)$ and $t(k)$, which are elements of the $S$ matrix.
We therefore refer to the equation $A(k)=0$ as the resonance equation.
In fact, the resonance equation has solutions not on the real axis but in the complex $k$ plane.

The above indirect definition of the resonant state consists of two steps: first, we find a scattering solution, assuming that $k$ is real;
then we do analytic continuation of $k$ and look for the zeros of $A(k)$ in the complex $k$ plane.
If we are interested only in finding resonant states, we can do it in one step;
we solve the Schr\"{o}dinger equation by putting $A\equiv0$ from the very beginning.
This constitutes the direct definition of the resonant state;
a resonant state is an eigenstate of the Schr\"{o}dinger equation under the condition $A\equiv0$, or that only outgoing waves $B$ and $C$ exist~\cite{Siegert39,Peierls59,Landau77,Hatano08}.
This last condition is called the Siegert condition.
We stress that the direct definition also gives the wave function of a resonant state.

For those who wonder if such solutions are possible, it is tutorial to solve a simple problem.
For example, consider the square-well potential:
$V(x)=-V_0$ for $|x|\leq a$ and $V(x)=0$ otherwise.
%\begin{align}\label{eq2-20}
%V(x)=\begin{cases}
%-V_0 & \mbox{for } |x|\leq a, \\
%0 & \mbox{otherwise.}
%\end{cases}
%\end{align}
We assume the wave function under the Siegert condition:
\begin{align}\label{eq2-30}
\psi(x)=\begin{cases}
Be^{-iKx} & \mbox{for }x<-a, \\
Fe^{iK'x}+Ge^{-iK'x} & \mbox{for }|x|\leq a, \\
Ce^{iKx} & \mbox{for }x>a
\end{cases}
\end{align}
with
\begin{align}\label{eq2-35}
K'\equiv \sqrt{K^2+\frac{2mV_0}{\hbar^2}}.
\end{align}
We use the uppercase letter $K$ for the wave numbers in order to emphasize that they are in general complex numbers.

\begin{figure}
\centering
\includegraphics[width=0.9\textwidth]{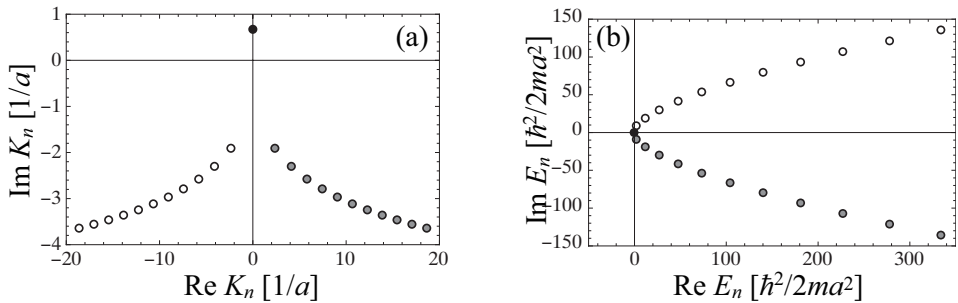}
\caption{The discrete solutions of Eqs.~(\ref{eq2-60}) and~(\ref{eq2-70}) with Eq.~(\ref{eq2-35}) 
in (a) the complex $K$ plane and (b) the complex $E$ plane.
We set $2mV_0 a^2/\hbar^2=1$.}
\label{fig2}
\end{figure}

We can simplify the problem by noting that the potential %~(\ref{eq2-20}) 
is an even function.
We can therefore classify all solutions into even and odd ones.
The even solutions must satisfy $B=C$ and $F=G$, which reduces~(\ref{eq2-30}) into
\begin{align}\label{eq2-40}
\psi(x)=\begin{cases}
2F\cos K'x & \mbox{for } 0\leq x \leq a, \\
Ce^{iKx} & \mbox{for }x>a.
\end{cases}
\end{align}
By requiring that $\psi(x)$ is continuous and differentiable at $x=a$, we have
%\begin{align}\label{eq2-50}
$2F\cos K'a=Ce^{iKa}$
%\qquad\mbox{and}\qquad
and $-2FK'\sin K'a=iCKe^{iKa}$,
%\end{align}
or
\begin{align}\label{eq2-60}
-K'\tan K'a = iK.
\end{align}
This equation is to be solved with the condition~(\ref{eq2-35}) 
and thereby yields discrete solutions.
Similarly for odd solutions, we have $B=-C$ and $F=-G$, and arrive at
\begin{align}\label{eq2-70}
%2iF\sin K'a=Ce^{iKa}
%2iFK'\cos K'a=iCKe^{iKa}
K'\cot K'a=iK,
\end{align}
which is again to be solved with the condition~(\ref{eq2-35}) 
and yields discrete solutions.
Figure~\ref{fig2} shows the discrete solutions thus obtained.
We will hereafter use the notations $K_n$, $E_n$ and $\psi_n(x)$ for any discrete solutions.

The solution on the positive imaginary $K$ axis, indicated by a solid circle in Fig.~\ref{fig2}(a), is actually a bound state.
We can see this immediately by putting $K_n=i\kappa_n$ with $\kappa_n>0$ in the assumed form of the wave function, Eq.~(\ref{eq2-30});
the wave function decays exponentially in space.
The corresponding eigenenergy is real and negative.
(In Fig.~\ref{fig2}(b), the solid circle is slightly on the negative side of the real axis.)

We refer to the solutions in the fourth quadrant of the complex $K$ plane, indicated by gray-filled circles in Fig.~\ref{fig2}(a), as resonant states, while the solutions in the third quadrant, indicated by empty circles, as anti-resonant states.
Each resonant state has a corresponding anti-resonant state with the same imaginary part and the real parts with the opposite signs.
In the complex $E$ plane, the resonant states are on the lower-half plane while the anti-resonant states are on the upper-half plane.
A resonant state and the corresponding anti-resonant state share a real part and have the opposite imaginary parts.
We can obtain the wave functions of the resonant and anti-resonant states by substituting $K$ in Eq.~(\ref{eq2-30}) with a complex value of $K_n$.

Finally, some systems can have solutions on the negative imaginary $K$ axis, generated when a resonant state and the corresponding anti-resonant state collide.
Such states, not present in the case of Fig.~\ref{fig2}, are referred to as anti-bound states~\cite{Ohanian74}.

\subsection{Non-hermiticity of open quantum systems}

We saw in the previous subsection that the standard Schr\"{o}dinger equation indeed can carry eigenstates with complex eigenvalues.
However, we all learn in our class of elementary quantum mechanics that the Hamiltonian
\begin{align}\label{eq2-100}
\hat{H}=\frac{\hbar^2}{2m}\frac{d^2}{dx^2}+V(x)
\end{align}
is a Hermitian operator (if $V(x)$ is a real function) and carries only real eigenvalues.
Why then did we find complex eigenvalues in the previous subsection?

In order to answer the question, let us review how we proved that $\hat{H}$ is a Hermitian operator.
We prepare an ``arbitrary" function $\phi(x)$;
for the moment, we do not specify which functional space this function $\phi(x)$ is in.
If the expectation value $\langle\phi|\hat{H}|\phi\rangle$ is real for an ``arbitrary" function $\phi(x)$, the Hamiltonian $\hat{H}$ is a Hermitian operator.
It is obvious that $\langle\phi|V|\phi\rangle$ is real.
We therefore check whether the expectation value of $d^2/dx^2$ is real or not.

For the moment, we define the expectation value as
\begin{align}\label{eq2-110}
\langle\phi|\frac{d^2}{dx^2}|\phi\rangle_L
=\int_{-L}^L \phi(x)^\ast  \phi''(x) dx,
\end{align}
where $[-L,L]$ is a region wide enough to contain the support of $V(x)$.
We would hope to take the limit $L\to\infty$ in the end, if possible.
Partial integration gives
\begin{align}\label{eq2-120}
\langle\phi|\frac{d^2}{dx^2}|\phi\rangle_L
=\left[\phi(x)^\ast\phi'(x)\right]_{x=-L}^L
-\int_{-L}^L\phi'(x)^\ast\phi'(x)dx.
\end{align}
The second term on the right-hand side is real.
The first term is real \textit{if} the function $\phi(x)$ is in the Hilbert space, or if $\lim_{L\to\infty}\phi(\pm L)=0$.
This is what we learned in our class of elementary quantum mechanics.
If $\phi(x)$ is not in the Hilbert space, the first term on the right-hand side of Eq.~(\ref{eq2-120}) can be indeed complex.

Let us then calculate the imaginary part of Eq.~(\ref{eq2-120}):
\begin{align}\label{eq2-130}
2i\mathop{\textrm{Im}}
\langle\phi|\frac{d^2}{dx^2}|\phi\rangle_L
=\left[\phi(x)^\ast\phi'(x)-\phi(x)\phi'(x)^\ast\right]_{x=-L}^L.
\end{align}
We can rewrite this in the following form:
\begin{align}\label{eq2-140}
2i\mathop{\textrm{Im}}
\langle\phi|\frac{d^2}{dx^2}|\phi\rangle_L
%&=\frac{i}{\hbar}\left[\phi(x)^\ast \hat{p} \phi(x)+\phi(x)(\hat{p}\phi(x))^\ast\right]_{x=-L}^L
%\nonumber\\
%&
=\frac{2i}{\hbar}\mathop{\textrm{Re}}\left[
\left.\phi(x)^\ast \hat{p} \phi(x)\right|_{x=L}
+\left.\phi(x)^\ast (-\hat{p}) \phi(x)\right|_{x=-L}
\right],
\end{align}
where $\hat{p}$ is the momentum operator $(\hbar/i)d/dx$.
In arbitrary number of dimensions, we can express the above in the form
\begin{align}\label{eq2-150}
\mathop{\textrm{Im}}\langle\phi|\hat{H}|\phi\rangle_\Omega
=-\frac{\hbar}{2m}\langle\phi|\hat{p}_\textrm{n}|\phi\rangle_{\partial\Omega},
\end{align}
where the left-hand side denotes the volume integral over a wide region $\Omega$, the right-hand side denotes the surface integral over $\partial\Omega$, and $\hat{p}_\textrm{n}$ is the normal component of the momentum on the surface.
(When we put $\phi$ to each resonant state $\psi_n$, the identity~(\ref{eq2-150}) reduces to the one derived previously~\cite{Berggren87,Moiseyev90,Masui99}.)

The equality~(\ref{eq2-150}) tells us that the Hamiltonian can be a non-Hermitian operator in a functional space that includes a function describing a state with momentum leak.
A state in which a momentum flux is leaking from an arbitrary wide region has a complex eigenenergy.
We can therefore claim that an open quantum system can generally have eigenstates with complex eigenvalues.
Let us now go back to the solutions marked in Fig.~\ref{fig2}.
When we use for $\phi$ the wave function of a resonant state or an anti-resonant state, the equality~(\ref{eq2-150}) confirms the relations
\begin{align}\label{eq2-155}
&\mathop{\textrm{Im}}E_n <0 \Longleftrightarrow \mathop{\textrm{Re}}K_n>0 \quad \mbox{for resonant states,}
\\
&\mathop{\textrm{Im}}E_n >0 \Longleftrightarrow \mathop{\textrm{Re}}K_n<0 \quad \mbox{for anti-resonant states.}
\end{align}

We can depict the situation more physically when we go into the time-dependent frame.
Using the time-dependent Schr\"{o}dinger equation
\begin{align}\label{eq2-160}
i\hbar\frac{d}{dt}\Phi(x,t)=\hat{H}\Phi(x,t),
\end{align}
we can prove another equality~\cite{Bohm89},
\begin{align}\label{eq2-170}
\frac{d}{dt}\langle\Phi|\Phi\rangle_\Omega=\frac{2}{\hbar}\mathop{\textrm{Im}}\langle\Phi|\hat{H}|\Phi\rangle_\Omega.
\end{align}
Combining Eqs.~(\ref{eq2-150}) and~(\ref{eq2-170}), we arrive at the equality
\begin{align}\label{eq2-180}
\frac{d}{dt}\langle\Phi|\Phi\rangle_\Omega=-\frac{1}{m}\langle\Phi|\hat{p}_\textrm{n}|\Phi\rangle_{\partial\Omega}.
\end{align}
This equality describes a clearly physical situation.
If a momentum flux is going out of a volume $\Omega$, the particle number in the volume decreases.
If a momentum flux is coming in, the particle number increases.
The former situation is a resonant state and the latter an anti-resonant state.
In this sense, the pair of a resonant and anti-resonant states spontaneously breaks the time-reversal symmetry~\cite{Petrosky88,Petrosky97a,Petrosky97b}.

\subsection{Particle-number conservation in a resonant state}

In the present subsection, we will show that the particle-number conservation is \textit{not} broken for a resonant state.
First, let us point out that the wave function~(\ref{eq2-30}) diverges in space for a resonant state because $\mathop{\textrm{Im}}K_n<0$.
The wave function of a resonant state, therefore, is not normalizable in the usual sense.
(A mathematical method of making the resonant wave function convergent has been devised and often called complex rotation~\cite{Aguilar71,Baslev71,Simon72,Moiseyev78,Simon79,McCurdy80,Moiseyev88,Csoto90,Moiseyev98,Ho83,Homma97,Myo97,Myo98,Masui99,Suzuki05,Aoyama06}.)
This is often seen as an unphysical feature of a resonant state.
In fact, we will show below that the spatial divergence is \textit{necessary} for the particle-number conservation for a resonant state, and hence is absolutely a physical feature.

In order to discuss the particle number conservation, we first count the number of particles in a wide region $[-L,L]$ as
\begin{align}\label{eq2-200}
N_L(t)=\int_{-L}^L\left|\Psi_n(x,t)\right|^2dx,
\end{align}
where $\Psi_n$ is the wave function of a resonant state with the time-dependent part:
\begin{align}\label{eq2-210}
\Psi_n(x,t)=\psi_n(x)e^{-iE_nt/\hbar}.
\end{align}
We would like to count the particle number after some time and see whether it changes or not.
At this point, we should remember the fact that particles are leaking from the boundaries in a resonant state.
The speed of the leak may be given by
%\begin{align}\label{eq2-220}
$v_n=(\hbar/m) \mathop{\textrm{Re}}K_n$.
%\end{align}
We therefore have to expand the integration region with the same speed in order to discuss the particle-number conservation correctly~\cite{Hatano08,Hatano09}.
This motivates us to replace Eq.~(\ref{eq2-200}) with
\begin{align}\label{eq2-230}
N_L(t)=\int_{-L(t)}^{L(t)}\left|\Psi_n(x,t)\right|^2dx
=e^{2t\mathop{\textrm{Im}}E_n/\hbar}\int_{-L(t)}^{L(t)}\left|\psi_n(x)\right|^2dx,
\end{align}
where the integration region $[-L(t),L(t)]$ expands as
%\begin{align}\label{eq2-240}
$L(t)=v_n t=(\hbar t/m) \mathop{\textrm{Re}}K_n$.
%\end{align}

Since $\psi_n(x)$ diverges for large $|x|$, the dominant contribution to the integration on the right-hand side of Eq.~(\ref{eq2-230}) comes from the portions around the bounds, $|x|\sim L(t)$.
We thereby have
\begin{align}\label{eq2-250}
N_L(t)&\simeq 2e^{2t\mathop{\textrm{Im}}E_n/\hbar}\int_{0}^{L(t)}e^{-2x\mathop{\textrm{Im}}K_n}dx
\nonumber\\
&\sim\exp\left(\frac{2t}{\hbar}\mathop{\textrm{Im}}E_n-2L(t)\mathop{\textrm{Im}}K_n\right)
\\ \label{eq2-255}
&=\exp\left[\frac{2t}{\hbar}\left(\mathop{\textrm{Im}}E_n-\frac{\hbar^2}{m}\mathop{\textrm{Re}}K_n\mathop{\textrm{Im}}K_n\right)\right]
\end{align}
The exponent vanishes because 
%\begin{align}\label{eq2-260}
$E_n=(\hbar^2/2m){K_n}^2$,
%\qquad\mbox{or}\qquad
or $\mathop{\textrm{Im}}E_n=(\hbar^2/m)\mathop{\textrm{Re}}K_n\mathop{\textrm{Im}}K_n$.
%\end{align}
Therefore the particle number is conserved in the above sense.
(We approximated the integral in the above, but we can show that the particle-number conservation holds exactly without approximation~\cite{Hatano09}.)
We note in Eq.~(\ref{eq2-250}) that the particle-number is conserved because the exponential decay in time exactly cancels the exponential divergence in space.
We therefore claim that the spatial divergence of the wave function of a resonant state is absolutely a physical feature.

We conclude this section by emphasizing that the resonant state given by its direct definition as an eigenstate of the Schr\"{o}dinger equation is, contrary to common misconception, a perfectly physical entity.
Several methods have been proposed to regularize the unnormalizable wave function by introducing seemingly unphysical definitions of the probability~\cite{Aguilar71,Baslev71,Simon72,Moiseyev78,Simon79,McCurdy80,Moiseyev88,Csoto90,Moiseyev98,Ho83,Homma97,Myo97,Myo98,Masui99,Suzuki05,Aoyama06,Zeldovich60,Hokkyo65,Romo68,Berggren70,Gyarmati71,Romo80,Berggren82,Berggren96,Madrid05}.
In contrast, we here used only the standard definition of the probability.
The above argument is a natural extension of the standard probability interpretation of the wave function to a resonant state.

\section{Resonant-state expansion of the Green's functions}
\label{sec3}

\subsection{Resonant-state expansion without back-ground integrals}

In the present section, we will review a recent proof of a resonant-state expansion of the sum of the retarded and advanced Green's functions.
The key point is that there is no background integral in the expansion.

\begin{figure}
\centering
\includegraphics[width=0.55\textwidth]{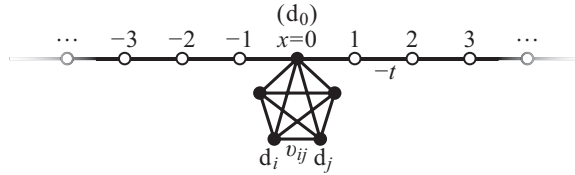}
\caption{A schematic view of the system~(\ref{eq3-10}), where a finite tight-binding scatterer indicated by the full circles is attached to an infinitely long tight-binding lead~\cite{Kittel86} indicated by the open circles.}
\label{fig3}
\end{figure}

The system that we consider is schematically shown in Fig.~\ref{fig3}.
The Hamiltonian is given in the general form
\begin{align}\label{eq3-10}
\hat{H}=\hat{H}_\textrm{dot}+\hat{H}_\textrm{lead},
\end{align}
where
\begin{align}\label{eq3-20}
\hat{H}_\textrm{dot}&=\sum_{i\in \textrm{dot}}\varepsilon_i |\textrm{d}_i\rangle\langle\textrm{d}_i|
+\sum_{\langle i,j \rangle}v_{ij}\left(|\textrm{d}_i\rangle\langle \textrm{d}_j|+|\textrm{d}_j\rangle\langle \textrm{d}_i|\right),
\\ \label{eq3-30}
\hat{H}_\textrm{lead}&=-t\sum_{x=-\infty}^\infty \left(|x+1\rangle\langle x|+|x\rangle\langle x+1|\right).
\end{align}
The first summation in Eq.~(\ref{eq3-20}) is taken over all sites $\{\textrm{d}_i\}$ in the scatterer, or namely the ``dot," and the second summation is taken over all pairs of the sites in the dot, while the summation in Eq.~(\ref{eq3-30}) is taken over the sites $\{x|x\in\mathbb{Z}\}$ in the lead.
Note that one of the sites in the scatterer, say $\textrm{d}_0$, is shared by the lead as $x=0$.
We may refer to this system as a generalized Friedrichs model~\cite{Sasada09}.
(We are trying to extend the following argument to more general systems\cite{Ordonez09}.)
%In the present paper, we do not consider any electron-electron interaction;
%it is a one-electron problem.

Because the lead is a tight-binding model, the distribution of the discrete eigenstates is modified in the following points.
First, the complex $K$ plane is limited to the Brillouin zone $-\pi\leq\mathop{\textrm{Re}}K\leq\pi$,
where the line $\mathop{\textrm{Re}}K=-\pi$ is identified with $\mathop{\textrm{Re}}K=\pi$.
Next, bound and anti-bound states can be also on the line $\mathop{\textrm{Re}}K=\pm\pi$.

For the above class of systems, we will focus on the Green's functions between two sites in the dot:
\begin{align}\label{eq3-40}
G_{ij}^{\textrm{R}/\textrm{A}}(E)=\langle \textrm{d}_i|\frac{1}{E-\hat{H}\pm i\delta}|\textrm{d}_j\rangle.
\end{align}
We will sketch a proof of the resonant state expansion of the form~\cite{Sasada09}
\begin{align}\label{eq3-50}
G_{ij}^\textrm{R}(E)+G_{ij}^\textrm{A}(E)=\sum_n\frac{\langle \textrm{d}_i| \psi_n\rangle\langle \tilde{\psi}_n|\textrm{d}_j\rangle}{E-E_n},
\end{align}
where the summation is taken over all discrete eigenstates, namely the bound states, the resonant states, the anti-resonant states and the anti-bound states.
We used the notation $\langle\tilde{\psi}_n|$ for the bra vector of a discrete state, because the bra vector, which will be obtained in Eq.~(\ref{eq3-65}) below, is not generally complex conjugate to the ket vector for resonant and anti-resonant states~\cite{Zeldovich60,Hokkyo65,Romo68,Berggren70,Gyarmati71,Romo80,Berggren82,Berggren96,Madrid05}.
We stress that there is no ``background" integral in the expansion~(\ref{eq3-50}).

The expansion~(\ref{eq3-50}) is a surprising result.
Normally, an expansion of a Green's function contains a background integral~\cite{Berggren70,GarciaCalderon76,Romo80,Berggren82,Berggren96,Madrid05,GarciaCalderon07a,GarciaCalderon07b}.
The point here is to \textit{add up} the retarded and the advanced Green's functions.
We will show below that the background integrals for the two Green's functions cancel out each other.
As a consequence, we will derive in the next section a formula for the electric conductance of the system~(\ref{eq3-10}) which has \textit{absolutely no} background integrals.
What seems to be a background integral of the conductance is actually the sum of the tails of all conductance peaks.

\subsection{Derivation of the resonant-state expansion}

Let us briefly describe the derivation of the expansion~(\ref{eq3-50}).
We start with the resolution of unity~\cite{Newton82}
\begin{align}\label{eq3-60}
\hat{1}=\sum_p|\psi_p^\textrm{b}\rangle\langle\psi_p^\textrm{b}|
+\int_{-\pi}^\pi\frac{dk}{2\pi}|\psi_k\rangle\langle\psi_k|,
\end{align}
where $\psi_p^\textrm{b}$ denotes a bound state and $\psi_k$ denotes a scattering state in continuum.
For later discussions, let us specify the scattering state $\psi_k$ in more details.
A scattering state may take the form
\begin{align}\label{eq3-61}
\psi_k(x)\equiv\langle x|\psi_k\rangle
%\nonumber\\
\propto\begin{cases}
A(k)e^{ikx}+B(k)e^{-ikx} & \mbox{for }x< 0, \\
C(k)e^{ikx} & \mbox{for }x>0.
\end{cases}
\end{align}
%where $\phi_k(0)=A(k)+B(k)=C(k)$.
We here symmetrize this and use the form
\begin{align}\label{eq3-62}
\psi_k(x)\propto%\begin{cases}
A(k)e^{-ik|x|}+\left(B(k)+C(k)\right)e^{ik|x|}. %& \mbox{for }x\leq 0, \\
%A(k)e^{-ikx}+\left(B(k)+C(k)\right)e^{ikx} & \mbox{for }x\geq 0.
%\end{cases}
\end{align}
(There are also anti-symmetrized solutions.
In the particular system~(\ref{eq3-10}), however, we have $\psi_k(0)=0$ for anti-symmetric solutions and hence the lead and the dot are completely decoupled.
Therefore, we focus on the symmetric solutions hereafter.)
The scattering state in Eq.~(\ref{eq3-62}) is then given by the normalized form
\begin{align}\label{eq3-63}
\psi_k(x)=e^{-ik|x|}+\frac{B(k)+C(k)}{A(k)}e^{ik|x|}.
\end{align}
Note that the scattering state has singularities at $A(k)=0$, which is nothing but the Siegert condition.
In fact, we can see that the residue at the singularity $k=K_n$
%\begin{align}\label{eq3-64}
%\psi_{K_n}(x)=
%\frac{B(K_n)+C(K_n)}{A'(K_n)}e^{iK_n|x|},
%\end{align}
%where $A'(k)\equiv dA(k)/dk$, 
is a properly normalized bound, resonant, anti-resonant, or anti-bound state:
\begin{align}\label{eq3-65}
|\psi_n\rangle\langle\tilde{\psi}_n|=\mathop{\textrm{Res}}_{k=K_n}|\psi_k\rangle\langle\psi_k|\equiv\lim_{k\to K_n}(k-K_n)|\psi_k\rangle\langle\psi_k|.
\end{align}

\begin{figure}
\hspace{0.1\textwidth}
\includegraphics[width=0.35\textwidth]{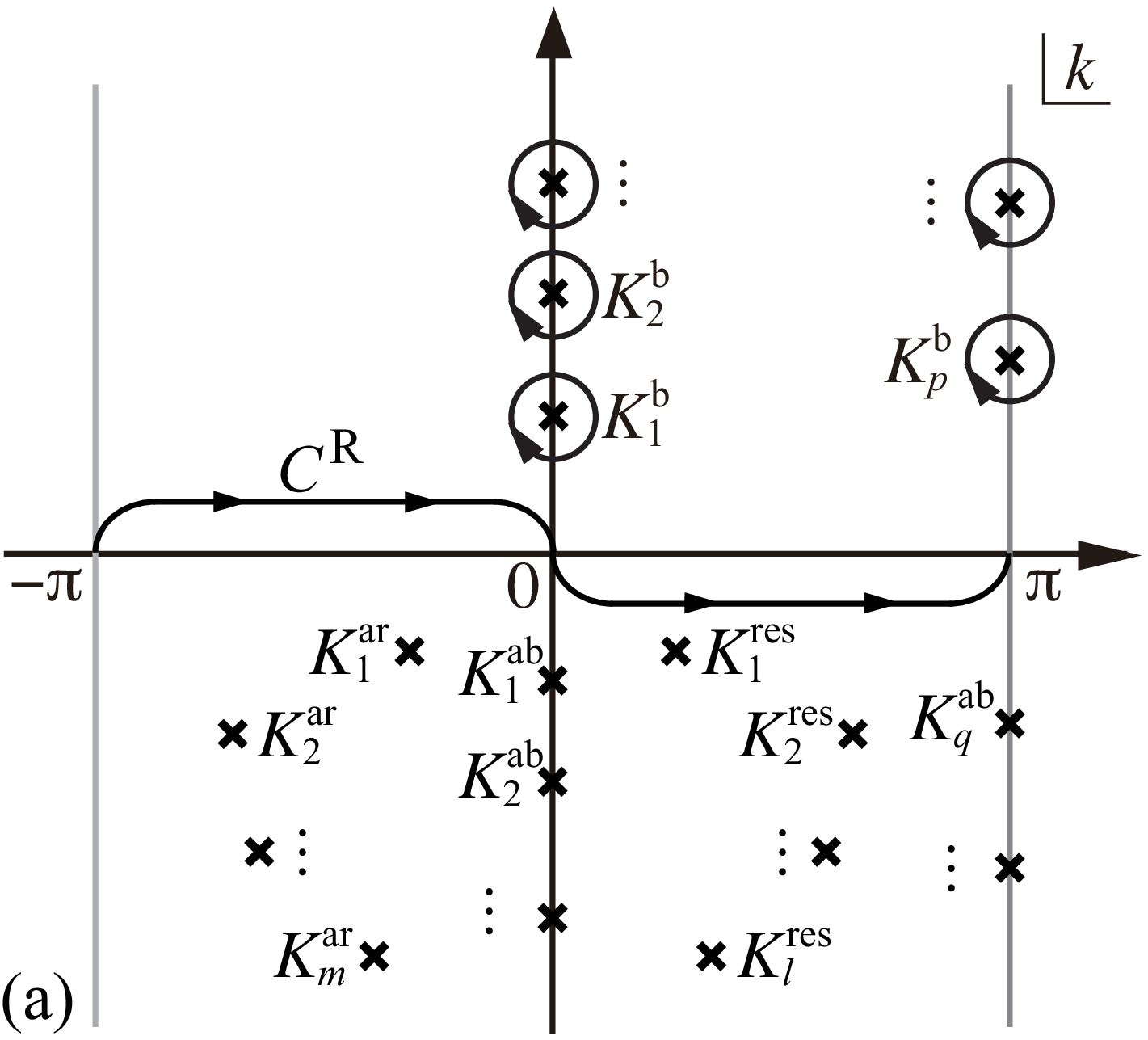}
\hfill
\includegraphics[width=0.35\textwidth]{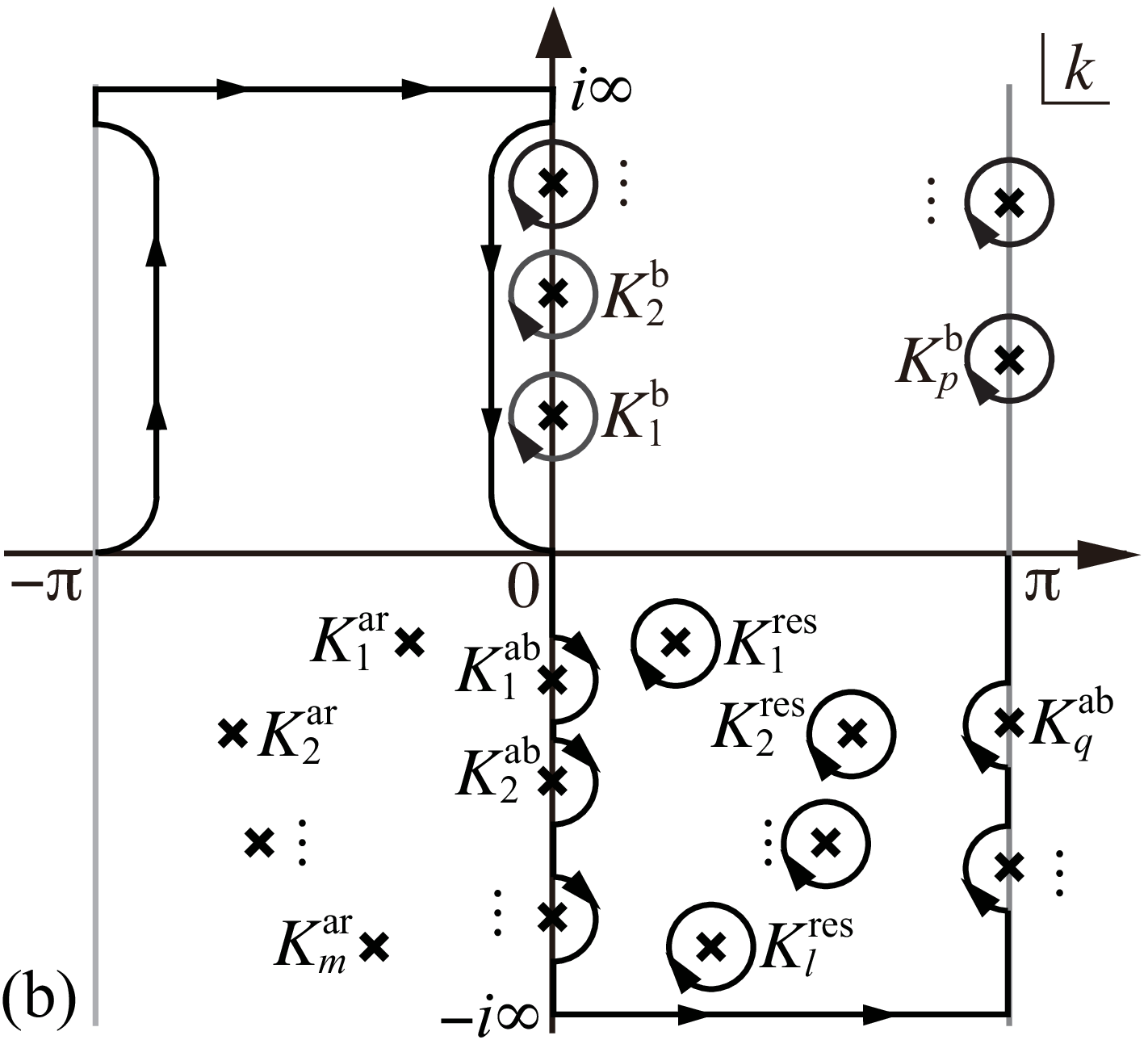}
\hspace{0.1\textwidth}

\caption{(a) The contour integration for the retarded Green's function in Eq.~(\ref{eq3-70}) and (b) the deformed contour.
Here $K_p^\textrm{b}$, $K_l^\textrm{res}$, $K_m^\textrm{ar}$ and $K_q^\textrm{ab}$ denote the singularities, respectively, for the bound states, the resonant states, the anti-resonant states and the anti-bound states.}
\label{fig4}
\end{figure}

We now go back to the resolution of unity~(\ref{eq3-60}). 
We thereby have
\begin{align}\label{eq3-70}
G_{ij}^{\textrm{R}/\textrm{A}}(E)=\sum_p\frac{\langle \textrm{d}_i|\psi_p^\textrm{b}\rangle\langle\psi_p^\textrm{b}|\textrm{d}_j\rangle}{E-E_p}
+\int_{C^{\textrm{R}/\textrm{A}}}\frac{dk}{2\pi}\frac{\langle \textrm{d}_i|\psi_k\rangle\langle\psi_k|\textrm{d}_j\rangle}{E-E_k},
\end{align}
where $E_p$ denotes the eigenenergy of the bound state $\psi_p^\textrm{b}$ and $E_k=-2t\cos k$ is the dispersion relation of the tight-binding lead~\cite{Kittel86}.
For the retarded Green's function, the integration contour $C^\textrm{R}$ is defined in Fig.~\ref{fig4}(a).
We then deform the integration contour as depicted in Fig.~\ref{fig4}(b);
we push down the contour on the positive real axis into negative imaginary infinity, while we push up the contour on the negative real axis into positive imaginary infinity.
In the process, we pick up the residues of various singularities.
We now remember Eq.~(\ref{eq3-65}), or that the residues are indeed contributions of bound, resonant, anti-resonant and anti-bound states.
Figure~\ref{fig4}(b) shows that, for the retarded Green's function, we pick up (i) the full contributions of the resonant states in the fourth quadrant, (ii) minus half of the contributions of the bound states on the positive imaginary axis as well as on the positive part of the line $\mathop{\textrm{Re}}K=-\pi$, and (iii) half of the contributions of the anti-bound states on the negative imaginary axis as well as on the negative part of the line $\mathop{\textrm{Re}}K=\pi$.

\begin{figure}
\hspace{0.1\textwidth}
\includegraphics[width=0.35\textwidth]{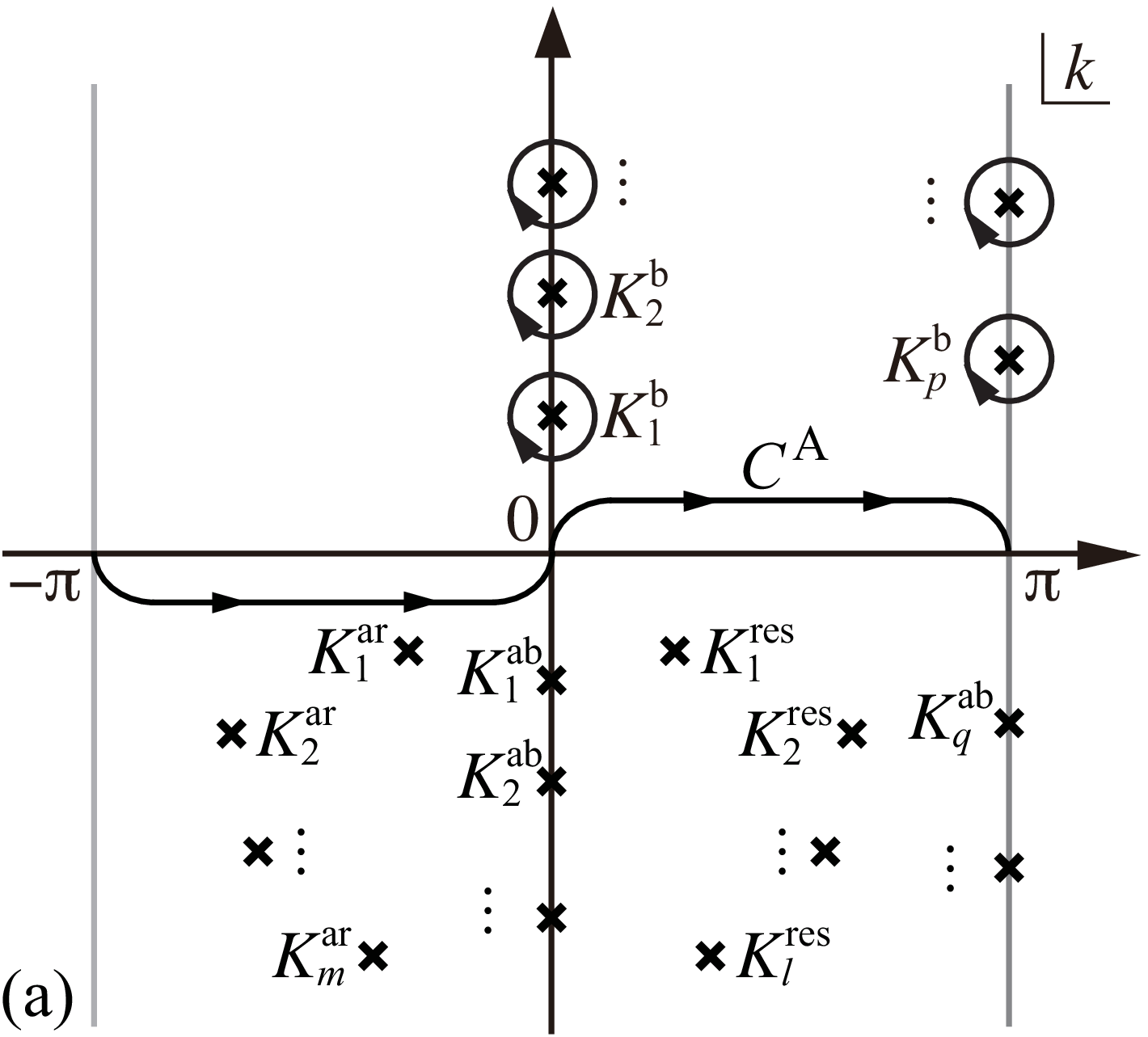}
\hfill
\includegraphics[width=0.35\textwidth]{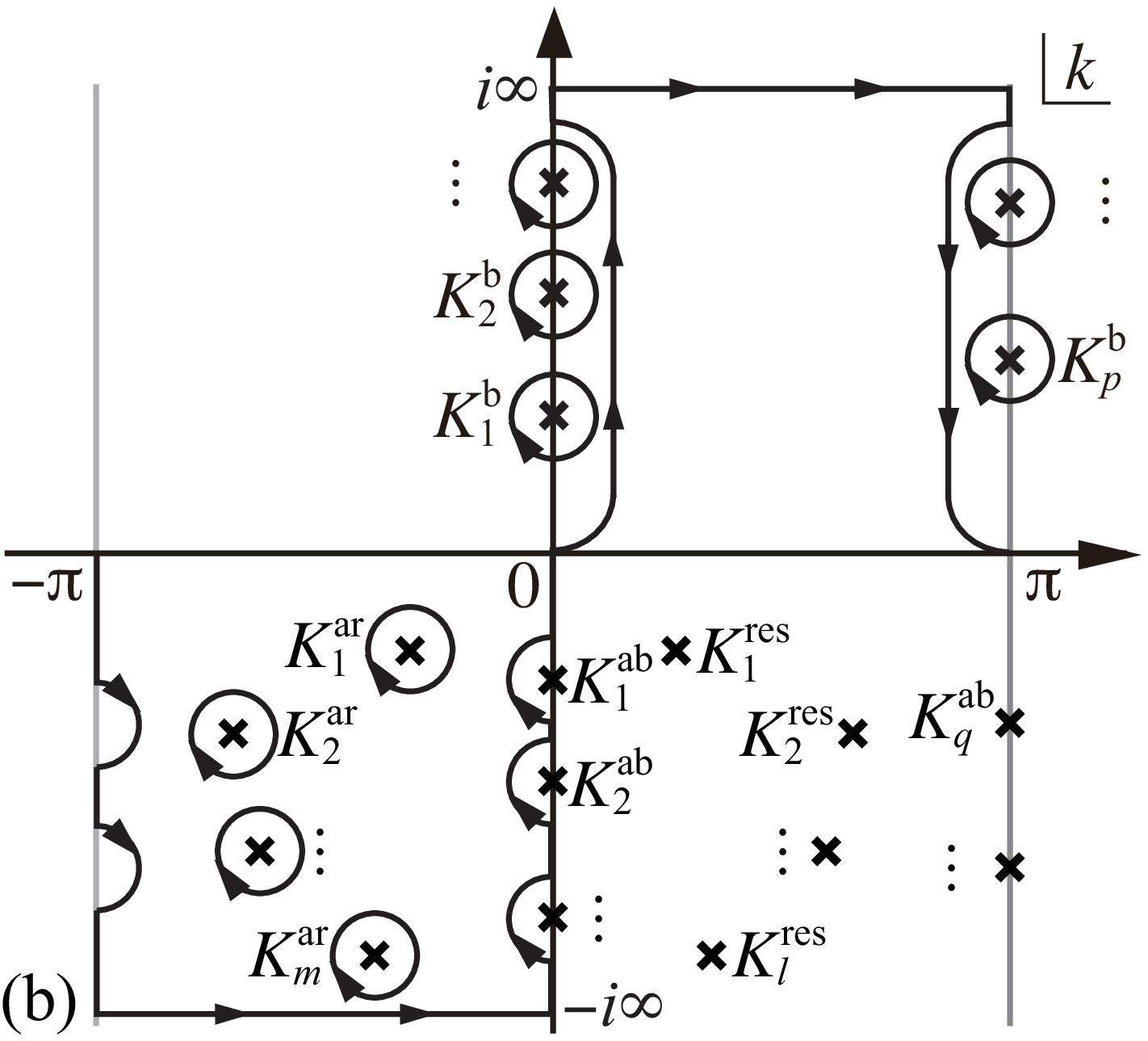}
\hspace{0.1\textwidth}

\caption{(a) The contour integration for the advanced Green's function in Eq.~(\ref{eq3-70}) and (b) the deformed contour.}
\label{fig5}
\end{figure}

For the advanced Green's function, the integration contour $C^\textrm{A}$ is defined in Fig.~\ref{fig5}(a).
This time we deform the integration contour as depicted in Fig.~\ref{fig5}(b);
we push up the contour on the positive real axis into positive imaginary infinity, while we push down the contour on the negative real axis into negative imaginary infinity.
This time, we pick up (i) the full contributions of the anti-resonant states in the third quadrant, (ii) the other minus half of the contributions of the bound states on the positive imaginary axis as well as on the positive part of the line $\mathop{\textrm{Re}}K=\pi$, and (iii) the other half of the contributions of the anti-bound states on the negative imaginary axis as well as on the negative part of the line $\mathop{\textrm{Re}}K=-\pi$.

\begin{figure}
\hspace{0.1\textwidth}
\includegraphics[width=0.35\textwidth]{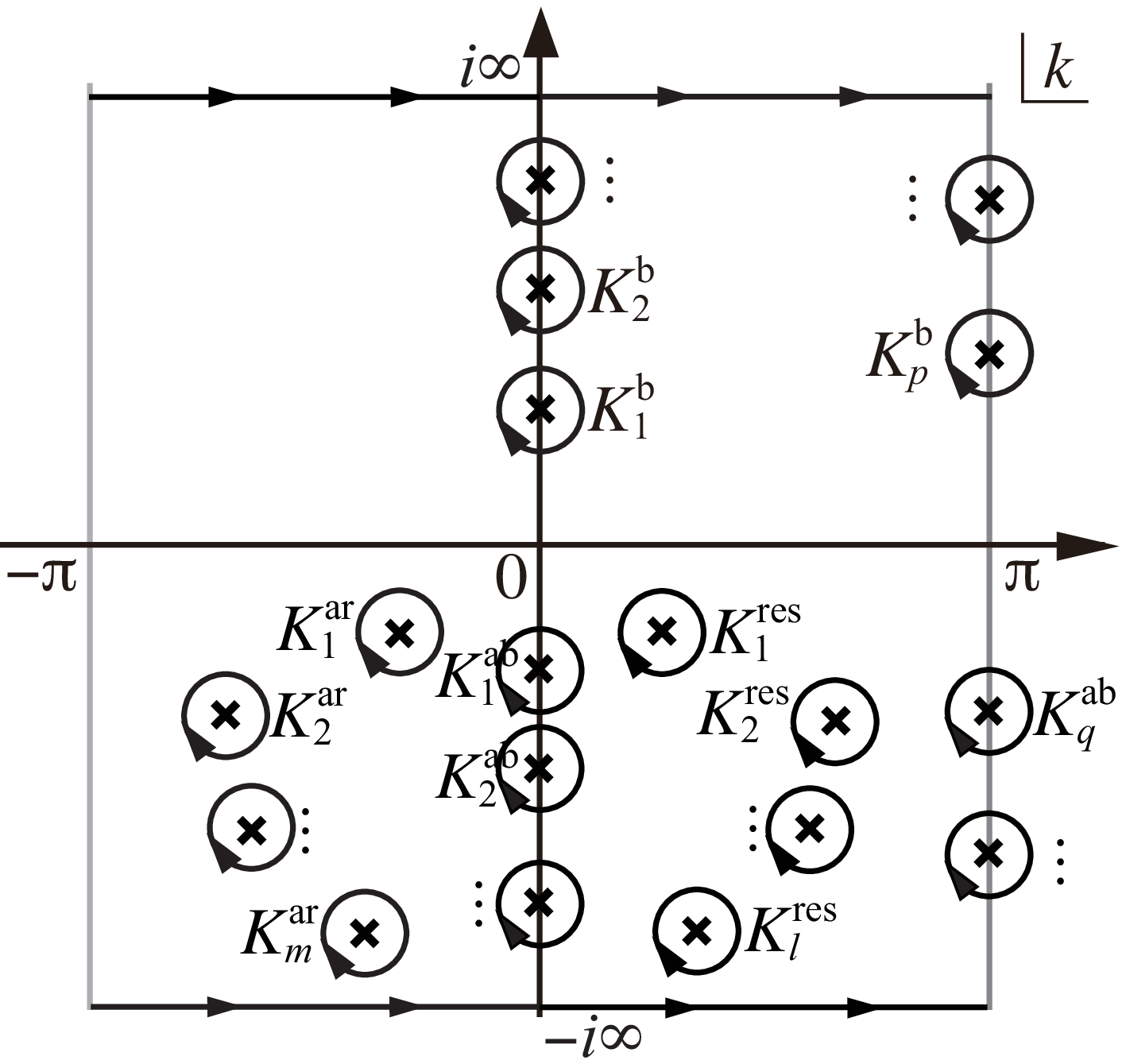}
\hfill
\includegraphics[width=0.35\textwidth]{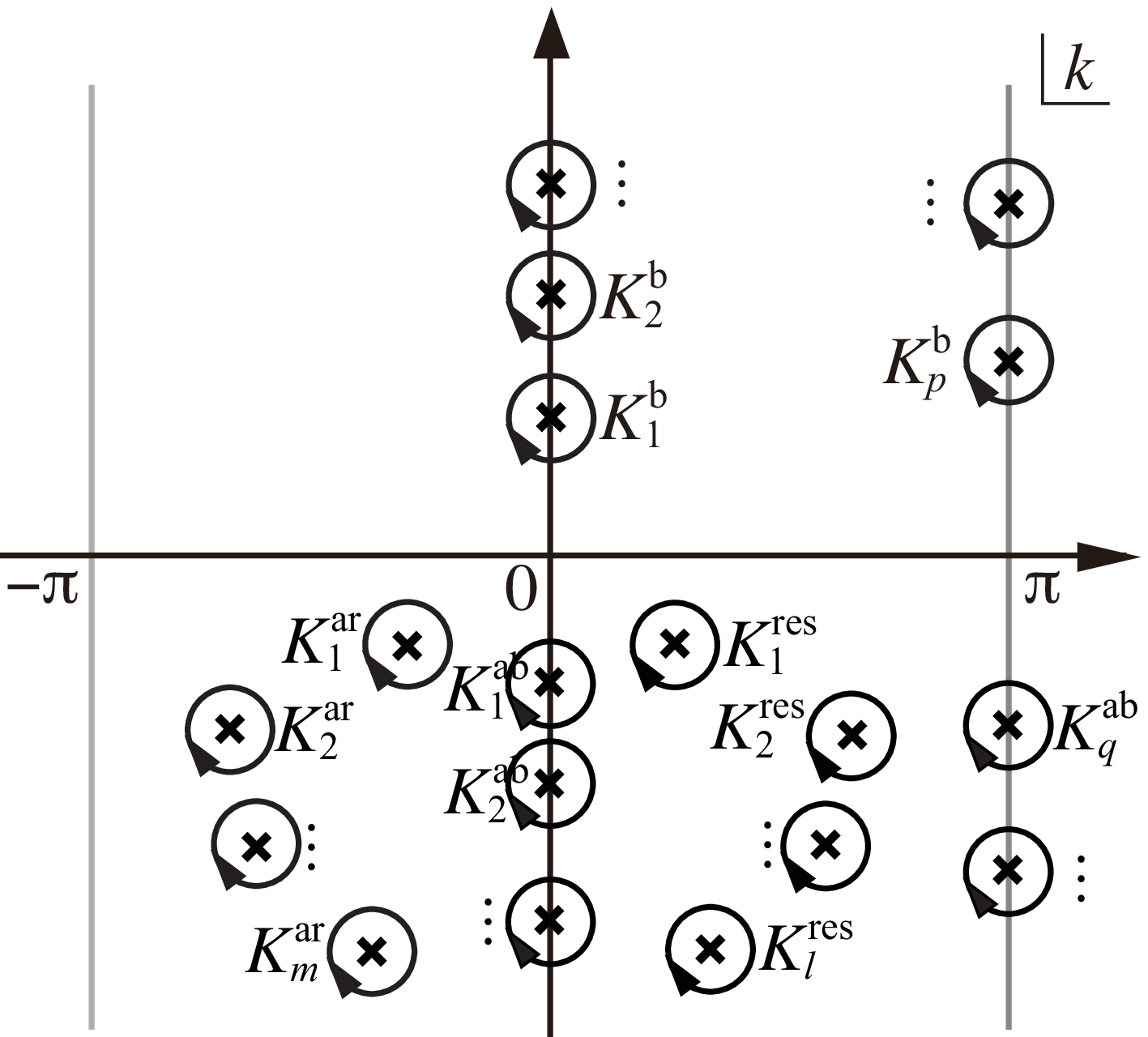}
\hspace{0.1\textwidth}

\caption{(a) The summation of the contour integrals in Fig.~\ref{fig4}(b) and Fig.~\ref{fig5}(b). (b) Only the residue integrals remain after the contour integrals in the infinities vanish.}
\label{fig6}
\end{figure}

By summing up both contours, we have the full contributions of all discrete eigenstates and the background contour integrals in the positive imaginary infinity as well as in the negative imaginary infinity; see Fig.~\ref{fig6}(a).
Note here the following two points.
First, the contour integrals along the imaginary axis as well as on the lines $\mathop{\textrm{Re}}K=\pm\pi$ completely cancel out each other.
Second, the contribution of the bound states are doubled by the first term on the right-hand side of Eq.~(\ref{eq3-70}), but the full contribution of the residue integral is subtracted.
Finally, we managed to prove that the ``background" integrals in the positive and negative imaginary infinities vanish~\cite{Sasada09}.
The equality~(\ref{eq3-50}) is thus proved as shown in Fig.~\ref{fig6}(b).

\section{Microscopic definition of the Fano parameters}
\label{sec4}

\subsection{Conductance formula without background integrals}

In the present section, using the expansion~(\ref{eq3-50}), we will derive the following formula for the conductance of the system~(\ref{eq3-10}):
\begin{align}\label{eq6-110}
\mathcal{G}(E)=\frac{2e^2}{h}\frac{1\pm\sqrt{1-\Omega(E)^2}}{2},
\end{align}
where
\begin{align}\label{eq6-90}
\Omega(E)\equiv\sqrt{4t^2-E^2}\sum_n\frac{\langle\textrm{d}_0|\psi_n\rangle\langle\tilde{\psi}_n|\textrm{d}_0\rangle}{E-E_n}.
\end{align}
It is again remarkable that the formula contains no background integrals.

Another point of the formula~(\ref{eq6-110}) is that it contains the square of the sum over the discrete eigenstates.
We therefore have interference between discrete eigenstates.
We will show that the interference generates asymmetry in the resonance peaks of the conductance.
The asymmetry is often referred to as the Fano asymmetry.
We will microscopically calculate the Fano parameters, which characterize the Fano asymmetry.
In fact, we find there are several types of the Fano asymmetry, one of which might have been missed in Fano's original argument~\cite{Fano61}.

For the proof of the formula~(\ref{eq6-110}), we start from the Landauer formula for the conductance of one-electron systems:
%\begin{align}\label{eq6-10}
$\mathcal{G}(E)=(2e^2/h)\left|t(E)\right|^2$,
%\end{align}
where $t(E)$ is the transmission amplitude~\cite{Datta95}.
We then use the Fisher-Lee relation for the transmission amplitude~\cite{Datta95}:
\begin{align}\label{eq6-20}
\mathcal{G}(E)=\frac{2e^2}{h}\mathop{\textrm{Tr}}\left(\Gamma(E)G^\textrm{R}(E)\Gamma(E)G^\textrm{A}(E)\right)
%\nonumber\\
=\frac{2e^2}{h}\times 4\left(4t^2-E^2\right)G^\textrm{R}_{00}(E)G^\textrm{A}_{00}(E)
\end{align}
where in the second equality we used the fact
%\begin{align}\label{eq6-30}
$\Gamma(E)=2\sqrt{4t^2-E^2}|\textrm{d}_0\rangle\langle\textrm{d}_0|$
%\end{align}
for the present system~(\ref{eq3-10}).

The $(0,0)$ elements of the retarded and advanced Green's functions are obtained as follows:
first, from the resonant-state expansion~(\ref{eq3-50}), we have
\begin{align}\label{eq6-40}
G^\textrm{R}_{00}(E)+G^\textrm{A}_{00}(E)=\Lambda_{00}\equiv\sum_n\frac{\langle\textrm{d}_0|\psi_n\rangle\langle\tilde{\psi}_n|\textrm{d}_0\rangle}{E-E_n};
\end{align}
second, we use the known identity
%\begin{align}\label{eq6-50}
$G^\textrm{A}(E)-G^\textrm{R}(E)=iG^\textrm{R}(E)\Gamma(E)G^\textrm{A}(E)$,
%\end{align}
or for the particular system of Eq.~(\ref{eq3-10}),
\begin{align}\label{eq6-60}
G^\textrm{A}_{00}(E)-G^\textrm{R}_{00}(E)=2i\sqrt{4t^2-E^2}G^\textrm{R}_{00}(E)G^\textrm{A}_{00}(E).
\end{align}
We thereby obtain
\begin{align}\label{eq6-70}
G^\textrm{R}_{00}(E)=\frac{\Lambda_{00}}{2}\left(1+\frac{1\pm\sqrt{1-\Omega(E)^2}}{i\Omega(E)}\right),
\quad
%\\ \label{eq6-80}
G^\textrm{A}_{00}(E)=\frac{\Lambda_{00}}{2}\left(1-\frac{1\pm\sqrt{1-\Omega(E)^2}}{i\Omega(E)}\right),
\end{align}
where we choose
\begin{align}\label{eq6-100}
\begin{cases}
\mbox{the upper signs ``$+$" if }
\left|\mathop{\textrm{Re}}G^\textrm{R}_{00}\right|<\left|\mathop{\textrm{Im}}G^\textrm{R}_{00}\right|,&
\\
\mbox{the lower signs ``$-$" if }
\left|\mathop{\textrm{Re}}G^\textrm{R}_{00}\right|>\left|\mathop{\textrm{Im}}G^\textrm{R}_{00}\right|.&
\end{cases}
\end{align}
%in Eqs.~(\ref{eq6-70}) and~(\ref{eq6-80}).
Using Eqs.~(\ref{eq6-70}) in Eq.~(\ref{eq6-20}), we arrive at the final expression~(\ref{eq6-110}).

In the following subsections, we will demonstrate that the crossing terms between discrete states contained in $\Omega(E)^2$ generate the Fano asymmetry of the resonance peaks in the conductance.
There are the following types of crossing terms:
\begin{enumerate}
\renewcommand{\labelenumi}{(\roman{enumi})}
\item (a resonant state) $\times$ (the corresponding anti-resonant state);
\item (a bound state) $\times$ (a resonant state $+$ the corresponding anti-resonant state);
\item (a resonant state $+$ the corresponding anti-resonant state)\\
$\times$ (another resonant state $+$ the corresponding anti-resonant state);
\item (a bound state) $\times$ (an anti-bound state);
\item (an anti-bound state)\\
$\times$ (a resonant state $+$ the corresponding anti-resonant state).
\end{enumerate}
Specifically, we will discuss the appearance of the crossing terms of the types (i), (ii) and (iii).
The crossing term of the type (i) will generate an asymmetry not discussed in Fano's original argument.

\subsection{T-shaped quantum dot: two types of Fano parameters}

\begin{figure}
\centering
\includegraphics[width=0.55\textwidth]{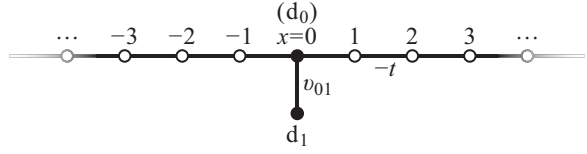}
\caption{A T-shaped quantum dot.}
\label{fig7}
\end{figure}

\begin{figure}
%\hspace{0.01\textwidth}
\includegraphics[width=0.5\textwidth]{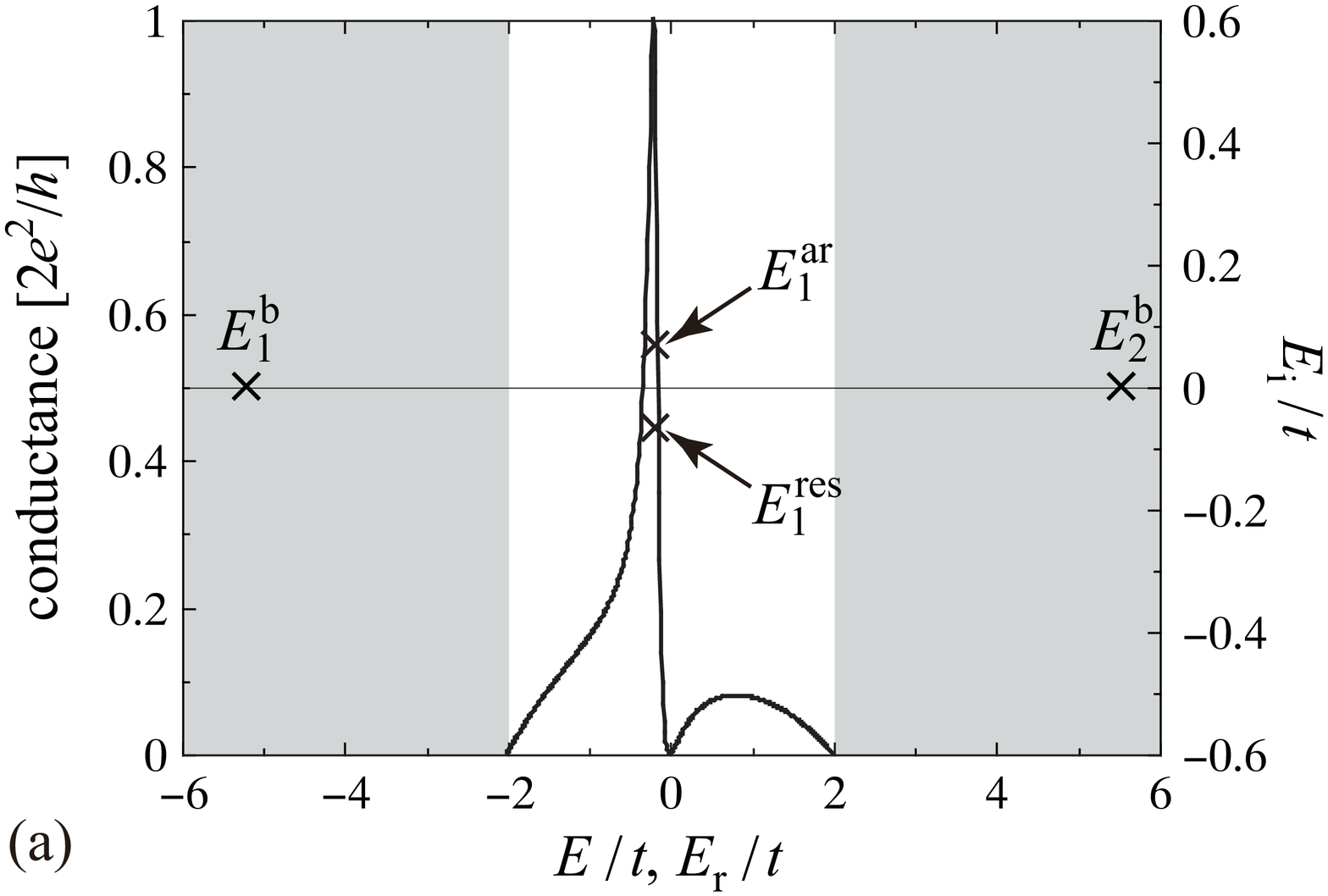}
%\hfill
\hspace{0.03\textwidth}
\includegraphics[width=0.46\textwidth]{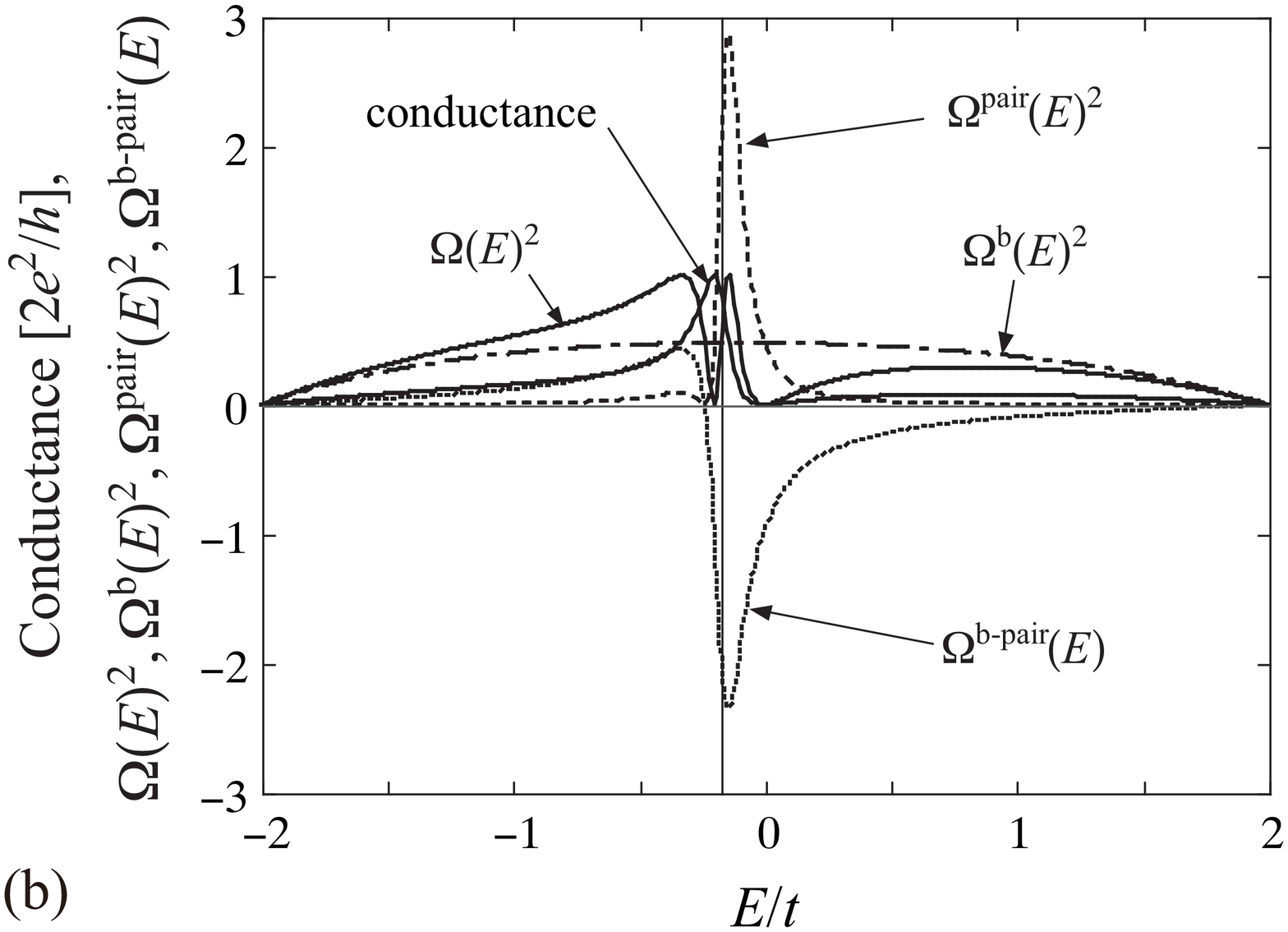}

\caption{(a) An asymmetric resonance peak of the conductance (the left axis) as well as the four discrete eigenvalues (the right axis) for $\varepsilon_0/t=5$, $\varepsilon_1/t=0$ and $v_{01}/t=v_{10}/t=1$.
(b) Various contributions to the conductance.
The gray vertical line indicates the real part of the resonant eigenvalue $\mathop{\textrm{Re}}E^\textrm{res}_1/t=-0.172712\ldots$.}
\label{fig8}
\end{figure}

In the present subsection, we consider the system shown in Fig.~\ref{fig7}.
The system accommodates two bound states $E^\textrm{b}_1$ and $E^\textrm{b}_2$, one resonant state $E^\textrm{res}_1$ and one anti-resonant state $E^\textrm{ar}_1$,
which are indicated by the crosses in Fig.~\ref{fig8}(a) for a specific parameter set.
In the same figure, we drew the conductance $\mathcal{G}(E)$.
The resonance peak around the real part of the resonant eigenvalue, $\mathop{\textrm{Re}}E^\textrm{res}_1$, exhibits a strong asymmetry.
We will dissect this asymmetry hereafter.

The term $\Omega(E)^2$ in the formula~(\ref{eq6-110}) contains
\begin{align}\label{eq6-200}
\frac{\Omega^\textrm{b}(E)^2}{4t^2-E^2}\equiv \left(\rho^\textrm{b}\right)^2,
\quad
\frac{\Omega^\textrm{pair}_1(E)^2}{4t^2-E^2}\equiv \left(\rho^\textrm{pair}_1\right)^2,
\quad
\frac{\Omega^\textrm{b-pair}_1(E)^2}{4t^2-E^2}\equiv 2\rho^\textrm{b}\rho^\textrm{pair}_1,
\end{align}
where
\begin{align}\label{eq6-210}
\rho^\textrm{b}\equiv
%\left(
\sum_{p=1,2}\frac{\langle\textrm{d}_0|\psi^\textrm{b}_p\rangle\langle\tilde{\psi}^\textrm{b}_p|\textrm{d}_0\rangle}{E-E^\textrm{b}_p},
%\right)^2,
%\\ \label{eq6-220}
\quad
\rho^\textrm{res}_1\equiv
%\left(
\frac{\langle\textrm{d}_0|\psi^\textrm{res}_1\rangle\langle\tilde{\psi}^\textrm{res}_1|\textrm{d}_0\rangle}{E-E^\textrm{res}_1}
+\frac{\langle\textrm{d}_0|\psi^\textrm{ar}_1\rangle\langle\tilde{\psi}^\textrm{ar}_1|\textrm{d}_0\rangle}{E-E^\textrm{ar}_1}.
%\right)^2,
%\\ \label{eq6-220}
%\left(
%\sum_{p=1,2}\frac{\langle\textrm{d}_0|\psi^\textrm{b}_p\rangle\langle\tilde{\psi}^\textrm{b}_p|\textrm{d}_0\rangle}{E-E^\textrm{b}_p}
%\right)
%\nonumber\\
%&\times\left(
%\frac{\langle\textrm{d}_0|\psi^\textrm{res}\rangle\langle\tilde{\psi}^\textrm{res}|\textrm{d}_0\rangle}{E-E^\textrm{res}}
%+\frac{\langle\textrm{d}_0|\psi^\textrm{ar}\rangle\langle\tilde{\psi}^\textrm{ar}|\textrm{d}_0\rangle}{E-E^\textrm{ar}}
%\right),
\end{align}
We plotted the terms in Eq.~(\ref{eq6-200}) in Fig.~\ref{fig8}(b) along with $\mathcal{G}(E)$ and $\Omega(E)^2$.
We can immediately see that the term $\Omega^\textrm{b}(E)^2$ does not have any asymmetry, but both the terms $\Omega^\textrm{pair}_1(E)^2$ and $\Omega^\textrm{b-pair}_1(E)^2$ do.

In order to derive a microscopic expression of the Fano parameter, we expand the terms $\Omega^\textrm{pair}_1(E)^2$ and $\Omega^\textrm{b-pair}_1(E)^2$ in the neighborhood of the real part of the resonant eigenenergy with respect to the normalized energy
%\begin{align}\label{eq6-230}
$\tilde{E}=(E-\mathop{\textrm{Re}}E^\textrm{res}_1)/\mathop{\textrm{Im}}E^\textrm{res}_1$.
%\end{align}
By using the expression
%\begin{align}\label{eq6-240}
$Ne^{i\theta}=\langle\textrm{d}_0|\psi^\textrm{res}_1\rangle\langle\tilde{\psi}^\textrm{res}_1|\textrm{d}_0\rangle$,
%\end{align}
we have
%\begin{align}\label{eq6-250}
%\Omega^\textrm{pair}(E)^2= (4t^2-E^2)
%\left(
%\frac{Ne^{i\theta}}{E-\mathop{\textrm{Re}}E^\textrm{res}-i\mathop{\textrm{Im}}E^\textrm{res}}
%+\textrm{c.c.}
%\right)^2,
%\end{align}
%which is followed by
\begin{align}\label{eq6-260}
\Omega^\textrm{pair}_1(E)^2\propto\left(\frac{q^\textrm{pair}_1+\tilde{E}}{1+\tilde{E}^2}\right)^2,
\end{align}
where the parameter
%\begin{align}\label{eq6-270}
$q^\textrm{pair}_1\equiv\tan\theta$
%\end{align}
%The parameter $q^\textrm{pair}_1$ 
controls the asymmetry of the term $\Omega^\textrm{pair}_1$;
the peak would be symmetric if $q^\textrm{pair}_1=0$.
Hence we may refer to it as the Fano parameter, although Eq.~(\ref{eq6-260}) is different from the original expression derived by Fano~\cite{Fano61}:
\begin{align}\label{eq6-280}
\mathcal{G}(E)\sim\frac{\left(q+\tilde{E}\right)^2}{1+\tilde{E}^2}.
\end{align}

On the other hand, we see that the term $\Omega^\textrm{b-pair}_1$ yields the asymmetry of Fano's original form~(\ref{eq6-280}) after some algebra.
We will not present the microscopic expression of the Fano parameter $q^\textrm{b-pair}$ here, but it is given in principle.
We suspect that Fano's original argument considered only the contribution of the term $\Omega^\textrm{b-pair}_1$.

\subsection{Three-site quantum dot: another type of Fano parameters}

\begin{figure}
\centering
\includegraphics[width=0.55\textwidth]{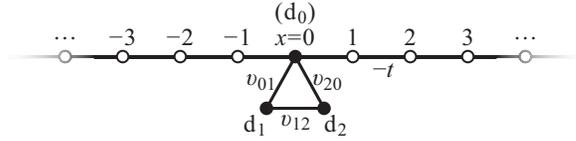}
\caption{A three-site quantum dot.}
\label{fig9}
\end{figure}

\begin{figure}
%\hspace{0.05\textwidth}
\includegraphics[width=0.5\textwidth]{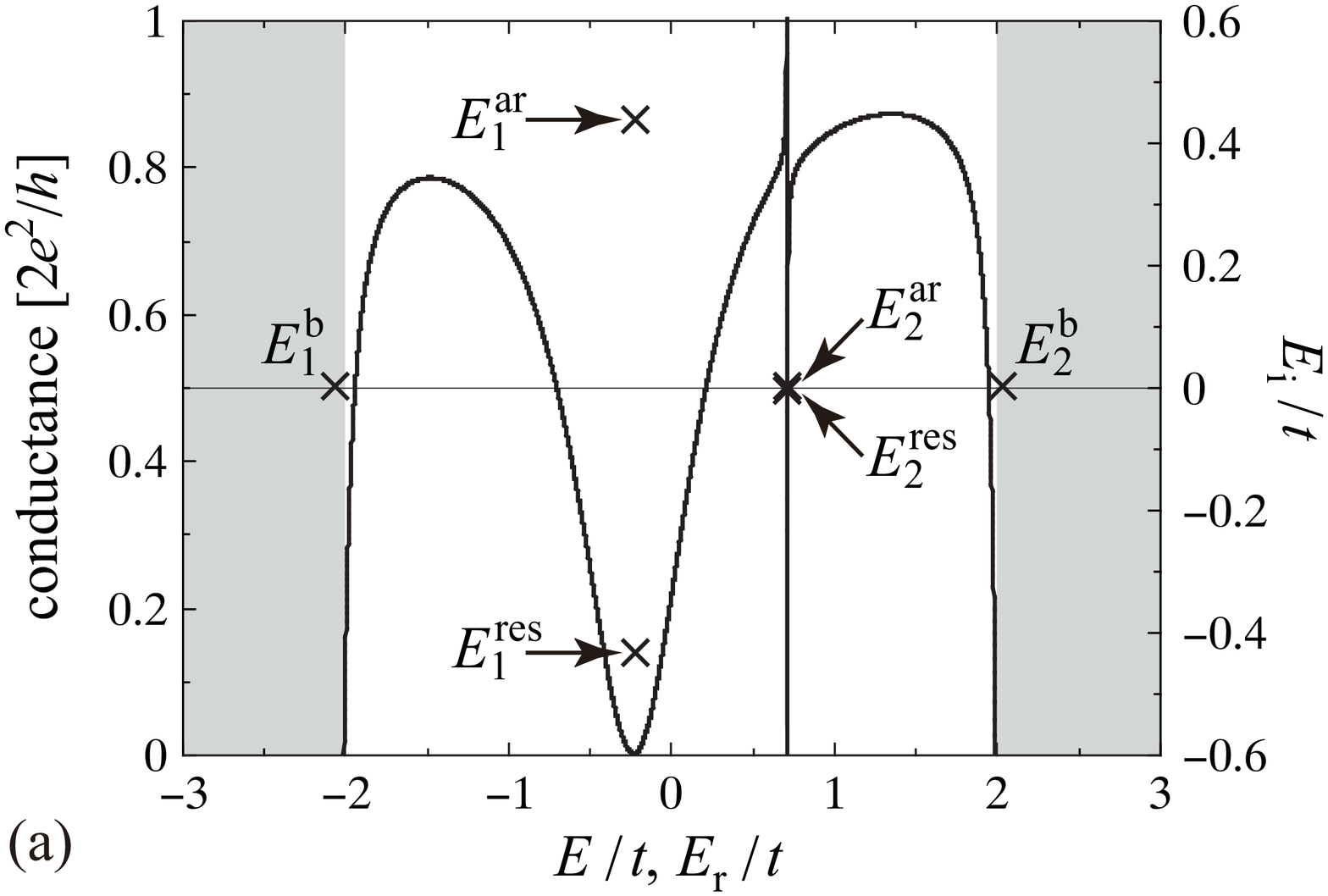}
%\hfill
\hspace{0.03\textwidth}
\includegraphics[width=0.46\textwidth]{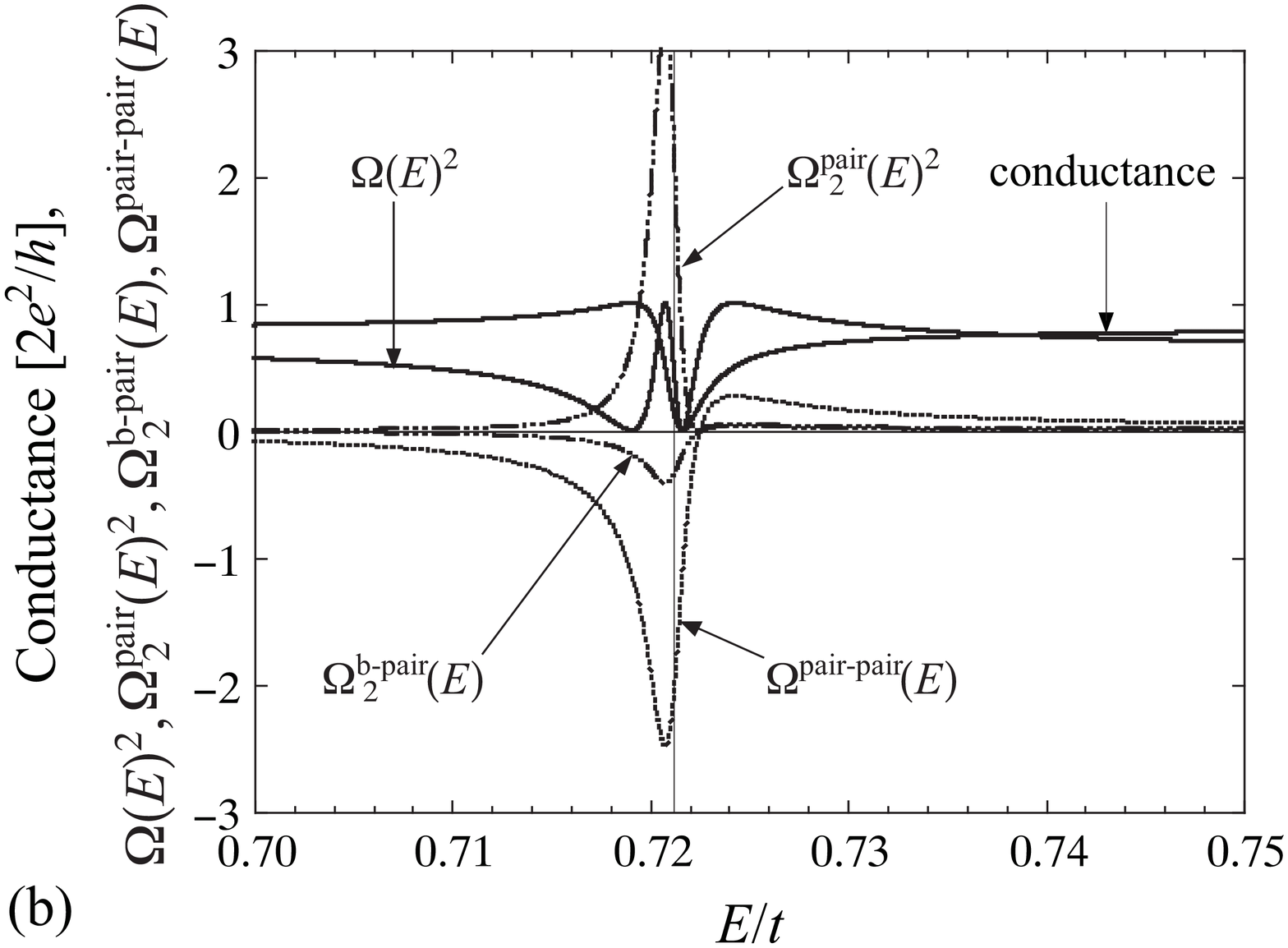}

\caption{(a) An asymmetric resonance peak of the conductance (the left axis) as well as the four discrete eigenvalues (the right axis) for $\varepsilon_0/t=0$, $\varepsilon_1/t=0$, $\varepsilon_2/t=0.5$, $v_{01}/t=v_{10}/t=0.8$, $v_{02}/t=v_{20}/t=0.5$ and $v_{12}/t=v_{21}/t=0.4$. 
(b) Various contributions to the conductance.
The vertical gray line indicates the real part of the second resonant eigenvalue $\mathop{\textrm{Re}}E^\textrm{res}_\textrm{2}/t=0.721170\ldots$.
}
\label{fig10}
\end{figure}

In the present subsection, we consider the system shown in Fig.~\ref{fig9}.
The system accommodates two bound states $E^\textrm{b}_1$ and $E^\textrm{b}_2$, two resonant states $E^\textrm{res}_1$ and $E^\textrm{res}_2$, and the corresponding two anti-resonant states $E^\textrm{ar}_1$ and $E^\textrm{ar}_2$,
which are indicated by the crosses in Fig.~\ref{fig10}(a) for a specific parameter set.
In the same figure, we drew the conductance $\mathcal{G}(E)$.
The resonance peak around the real part of the second resonant eigenvalue, $\mathop{\textrm{Re}}E^\textrm{res}_2$, exhibits a strong asymmetry.

The term $\Omega(E)^2$ for this system can be dissected into the terms in Eq.~(\ref{eq6-200}) as well as
\begin{align}\label{eq6-300}
\frac{\Omega^\textrm{pair}_2(E)^2}{4t^2-E^2}\equiv\left(\rho^\textrm{pair}_2\right)^2,
\quad
\frac{\Omega^\textrm{b-pair}_2(E)^2}{4t^2-E^2}\equiv2\rho^\textrm{b}\rho^\textrm{pair}_2,
\quad
\frac{\Omega^\textrm{pair-pair}(E)^2}{4t^2-E^2}\equiv2\rho^\textrm{pair}_1\rho^\textrm{pair}_2,
\end{align}
which we plotted in Fig.~\ref{fig10}(b) along with $\mathcal{G}(E)$ and $\Omega(E)^2$.
The term $\Omega^\textrm{b}(E)^2$ does not have any asymmetry, but the other terms do.

Expanding these terms in the neighborhood of the real part of the second resonant eigenenergy with respect to the normalized energy
%\begin{align}\label{eq6-340}
$\tilde{E}=(E-\mathop{\textrm{Re}}E^\textrm{res}_2)/\mathop{\textrm{Im}}E^\textrm{res}_2$,
%\end{align}
we can derive microscopic expressions for the Fano parameters $q^\textrm{pair}_2$, $q^\textrm{b-pair}_2$, and $q^\textrm{pair-pair}$.
In this case, we found that $q^\textrm{pair-pair}$ is relatively large, particularly when the first resonant eigenenergy $E^\textrm{res}_1$ approaches the second resonant eigenenergy $E^\textrm{res}_2$.
Indeed, we can generally argue that, when a resonant state with a large imaginary part is beside a resonant state with a small imaginary part, a strong asymmetry develops around the real part of the latter~\cite{Sasada05}.
This implies that even a resonant state with a large imaginary part can manifest its effect in an observable, contrary to the conventional conception that a resonant state far from the real axis is negligible.

\section{Summary}

In the present paper, we have reviewed two of our recent studies.
First, we considered physical aspects of the Siegert boundary condition.
By deriving the equality~(\ref{eq2-150}) and thereby the equality~(\ref{eq2-180}), we showed that in a resonant state as an eigenstate of the Schr\"{o}dinger equation, particles in the central region leak into infinities and hence the number of particles decay in time.
We also showed that an anti-resonant state is the time reversal of the corresponding resonant state.
Using the above picture of the resonant state, we argued that the particle number is conserved even in a resonant state, and hence the probabilistic interpretation of quantum mechanics persists.

Second, we derived a resonant-state expansion of the sum of the retarded and advanced Green's functions.
It is a key aspect that the expansion did not contain any background integrals.
We presented a brief sketch of the proof of the expansion for a specific class of systems.
We are now trying to generalize the proof~\cite{Ordonez09}.

Finally, we considered the Fano asymmetry of resonant peaks in the conductance profile.
We argued that the asymmetry can be traced back to interference between various discrete eigenstates.
In particular, we suggested that the interference between a resonant and the corresponding anti-resonant states were not considered in Fano's original, somewhat phenomenological argument.

\section*{Acknowledgements}
The present author appreciates great benefit from Yukawa International Program for Quark-Hadron Sciences (YIPQS).
He expresses sincere gratitude to Prof.~T.~Petrosky, Prof.~G.~Ordonez, Prof.~H.~Nakamura and Dr.~K.~Sasada for collaborating on these topics.
Helpful discussions with Dr.~A.~Nishino and Dr.~T.~Imamura are gratefully acknowledged.
The present study is supported by a CREST-JST project.
%We would like to thank ...........

%\appendix
%\section{First Appendix} %Empty argument \section{} yields `Appendix'. 
%
%\section{Second Appendix}

\end{document}